\documentclass[preprint2]{aastex62}
\usepackage{color}
\usepackage{url}
\usepackage{bm}

\received{MMMM DD, YYYY}
\revised{\today}
\accepted{MMMM DD, YYYY}
\submitjournal{ApJ}
\shorttitle{The Role of a Tiny Brightening for Huge Geo-effective Solar Eruption}
\shortauthors{Bamba et al.}

\begin{document}

\title{The Role of a Tiny Brightening in a Huge Geo-effective\\Solar Eruption Leading to the St Patrick's Day Storm}

\correspondingauthor{Yumi Bamba}
\email{y-bamba@nagoya-u.jp}

\author{Yumi Bamba}
\affiliation{Institute of Space and Astronautical Science (ISAS)/Japan Aerospace Exploration Agency (JAXA)\\ 3-1-1 Yoshinodai, Chuo-ku, Sagamihara, Kanagawa 252-5210, Japan}

\author{Satoshi Inoue}
\affiliation{Institute for Space-Earth Environmental Research (ISEE)/Nagoya University\\Furo-cho, Chikusa-ku, Nagoya, Aichi 464-8601, Japan}

\author{Keiji Hayashi}
\affiliation{Key Laboratory of Solar Activity, National Astronomical Observatories, Chinese Academy of Sciences, Beijing 100012, China}
\affiliation{State Key Laboratory for Space Weather, National Space Science Center, Chinese Academy of Sciences, Beijing 100190, China}
\affiliation{Institute for Space-Earth Environmental Research (ISEE)/Nagoya University, Furo-cho, Chikusa-ku, Nagoya, Aichi 464-8601, Japan}
\nocollaboration

\begin{abstract} 

The largest magnetic storm in solar cycle 24 was caused by a fast coronal mass ejection (CME) that was related to a small C9.1 flare that occurred on 15 March 2015 in solar active region (AR) NOAA 12297. The purpose of this study is to understand the onset mechanism of the geo-effective huge solar eruption. We focused on the C2.4 flare that occurred prior to the C9.1 flare of the filament eruption. The magnetic field structure in the AR was complicated: there were several filaments including the one that erupted and caused the CME. We hence carefully investigated the photospheric magnetic field, brightenings observed in the solar atmosphere, and the three-dimensional coronal magnetic field extrapolated from nonlinear force-free field modeling, using data from {\it Hinode} and {\it Solar Dynamics Observatory}. We found three intriguing points :  (1) There was a compact but noticeably highly twisted magnetic field structure that is represented by a small filament in the C2.4 flaring region, where a tiny precursor brightening was observed before the C2.4 flare. (2) The C2.4 flaring region is located in the vicinity of a foot point of the closed field that prohibits the filament from erupting. (3) The filament shows a sudden eruption after the C2.4 flare and accompanying small filament eruption. From our analysis, we suggest that a small magnetic disturbance that was represented by the tiny precursor brightening at the time of the C2.4 flare is related to the trigger of the huge filament eruption.
\end{abstract}



\section{Introduction} \label{sec:intro}

Solar eruptions, such as flares, coronal mass ejections (CMEs), and filament eruptions, are the biggest explosions in the solar system.
They are driven by magnetic energy stored in the solar corona.
Sunspot evolution and/or photospheric motions often form a twisted magnetic flux rope, which erupts into interplanetary space with a flare and CME.
The magnetic flux rope is sometimes represented by a filament \citep[e.g.][]{labrosse10, mackay10}, which is dense and cool plasma residing in the hot corona.
Recent high resolution H$\alpha$ observations show the twisted structure of flux ropes \citep[e.g.][]{wang15, xue16}.
The magnetic flux rope can disturb the space environment around the Earth when it reaches the Earth having a south-oriented magnetic field.
A CME, which contains the magnetic flux rope, reaches the Earth within 3-4 days, but it can take approximately 14 hours at the shortest \citep{knipp18}.
Therefore, it is important to understand the onset mechanism(s) of solar flux rope eruption not only from the solar physics perspective but also for space weather forecast.

There are many observational and numerical studies that attempt to reveal the onset mechanism(s) of solar flux rope eruptions.
For instance, several models are proposed on the basis of ideal or non-ideal magnetohydrodynamics (MHD) models.
Those models emphasize different aspects of the mechanism of solar eruptions: formation of magnetic flux rope through magnetic reconnection and evolution of the flux rope to eruption by the MHD instability.
The torus instability \citep[e.g.][]{bateman78, kliemtorok06, torokkliem07, demoulin10, olmedo_zhang10, kliem14}, helical kink instability \citep[e.g.][]{gerrard01, torok04, fangibson03, torokkliem05}, tether-cutting reconnection \citep{moore01, moore06, ishiguro_kusano17}, and magnetic break-out \citep{antiochos99, karpen12} have been proposed as physical processes leading to solar eruptions so far.
However, it is still not clear what triggers an initiation of the solar flux rope eruption.

It is statistically known that the incidence of flux rope eruption, which is represented by a CME and/or filament eruption, correlates with the X-ray peak flux of the corresponding flare \citep{yashiro06}.
Huge numbers of studies relate the onset processes of flux rope eruption to large flares \citep[e.g.][]{inoue15, Woods17, bamba13, bamba17a, bamba17b, zuccarello17, yan17}.
However, not only large flares but also small flares relate to flux rope eruption \citep[e.g.][]{louis15, dudik17}, and even small flares can have a geo-effective impact.

AR NOAA 12297 appeared on the southeast solar limb on March 6, 2015, and produced large numbers of flares during its disk passage.
The largest flare in terms of {\it GOES}\footnote{\it{Geostationary Operational Environmental Satellite}} soft X-ray class was the X2.1 flare on March 11.
In the present study, we focused on the sequential C2.4 and C9.1 flares that occurred on March 15, when the AR was located at S22$^{\circ}$ W25$^{\circ}$.
The C-class flares seem to relate to a fast CME that became one of the causes of the largest magnetic storm so far during solar cycle 24, which occurred on March 17, 2015, with a minimum {\it Dst} index of -223 nT \citep[e.g.,][]{liu15, kataoka15}.

In this study, we aim to answer the question: why such a huge solar eruption occurred associated with a small C-class flare?
In order to answer the question, we analyzed the dynamics of the filament before and after its eruption using extreme-ultraviolet (EUV) images.
Moreover, we investigated the photospheric magnetic field structures close to a foot point of the closed field that prohibited the filament erupting.
We compared the results from the data analysis with the extrapolated three-dimensional coronal magnetic field structures obtained from a nonlinear force-free field (NLFFF) model.
As a result, we identified a tiny brightening that played an important role in triggering the huge geo-effective filament eruption.

The paper is organized in the following manner.
We first review the C-class flares and source AR, which we focused on in this study, and describe the data sets, analysis methods, and the NLFFF modeling in Section~\ref{sec:analysis}.
Then we report the results from the data analysis in Section~\ref{sec:results}.
In Section~\ref{sec:discussion}, we discuss the onset process of the huge filament eruption by comparing the analysis results and the NLFFF modeling.
Finally, we summarize the present study in Section~\ref{sec:summary}.

\section{Data Analysis and Modeling} \label{sec:analysis}

We used {\it Solar Dynamics Observatory} \citep[{\it SDO};][]{pesnell12} data to investigate the temporal changes of the erupting filament and surrounding global features in the photosphere and chromosphere and to extrapolate three-dimensional coronal magnetic field.
Moreover, we analyzed {\it Hinode} \citep{kosugi07} data to precisely investigate the photospheric magnetic field and evolution of chromospheric brightenings.

\subsection{Solar Dynamics Observatory} \label{sec:analysis_sdo}

We employed data obtained by {\it SDO} to investigate the global structure containing multiple filament channels around the AR.
The Atmospheric Imaging Assembly \citep[AIA;][]{lemen12} images the solar atmosphere in 10 wavelengths.
The spatial sampling and spatial resolution were $0.6^{\prime\prime}$/pixel and $1.5^{\prime\prime}$, respectively.
We analyzed AIA/304 \AA (\ion{He}{2} line) and 93 \AA (\ion{Fe}{12} and \ion{Fe}{24} lines) images \footnote{``{\tt aia.lev1\_euv\_12s}'' series downloaded from Joint Science Operations Center (JSOC) at Stanford University (\url{http://jsoc.stanford.edu/})}.
The former is sensitive to emission from the chromosphere and transition region ($\log T \sim 4.7$) and the latter is sensitive to the corona and hot flare plasma ($\log T \sim 6.2, 7.3$).
The data coverage in our analysis was from 18 UT on March 14 to 06 UT on March 15, 2015, and the cadence was 12 seconds.
AIA images were calibrated by applying the {\tt aia\_prep} procedure of the {\it Solar Soft-Ware (SSW)} package, and we investigated the topological change of the erupting filament.

In addition, we analyzed vector magnetograms obtained by the Helioseismic and Magnetic Imager \citep[HMI;][]{schou12} in the \ion{Fe}{1} line (6173 \AA).
The vector magnetograms have been released as SHARP \citep[Spaceweather HMI Active Region Patch;][]{bobra14} data series \footnote{``{\tt hmi.sharp\_cea\_720s}'' series}, in which the magnetic field has been derived assuming the Milne-Eddington atmosphere and remapped to a Lambert Cylindrical Equal-area (CEA) projection.
The $180^{\circ}$ ambiguity in the horizontal component was resolved using a minimum energy method \citep{metcalf94}.
The spatial sampling and spatial resolution were $0.5^{\prime\prime}$/pixel and $0.9^{\prime\prime}$, respectively.
We used a single SHARP image taken at 00 UT on March 15, 2015, for our NLFFF modeling (see Section~\ref{sec:analysis_nlfff} for details).
Moreover, we extracted an HMI vector magnetogram at 00 UT on March 15 for a larger region than the SHARP field-of-view (FOV) around the AR in order to extrapolate the three-dimensional (3D) coronal magnetic field (see \ref{sec:analysis_nlfff} for details).
We cut out a region centered around $195^{\circ}$ Carrington Longitude and $20^{\circ}$ latitude south from the solar equator.
We first converted the magnetic field vector data in CCD coordinates $({B}_{x}, {B}_{y}, {B}_{z})$ \footnote{``{\tt hmi.B\_720s[2015.03.15\_00:00:00\_TAI]}'' at the JSOC database} to those in the spherical coordinate system $({B}_{r}, {B}_{\theta}, {B}_{\phi})$.
Then we projected the local orthogonal vector components onto the $800 \times 800$ pixel CEA map \footnote{We used the {\tt mtrack} code, available at the JSOC, in the CEA projection step to create maps in the frame moving with the solar differential rotation.} with $0.03^{\circ} \times 0.03^{\circ}$ heliocentric angle resolution at the map center.

\subsection{Hinode} \label{sec:analysis_hinode}

We analyzed chromospheric images obtained by {\it Hinode}/Solar Optical Telescope \citep[SOT; ][]{tsuneta08} in the \ion{Ca}{2} H line (3968 {\AA}).
The data covered 12 hours from 18 UT on March 14 to 06 UT on March 15, 2015, with 1 minute cadence and a $183^{\prime\prime} \times 108^{\prime\prime}$ FOV.
Continuum images in G-band (\ion{CH}{1} line, 4305 \AA) were also taken every 10 minutes with the same FOV.
The spatial sampling was $0.108^{\prime\prime}$/pixel by $2 \times 2$ summing and the spatial resolution was approximately $0.2^{\prime\prime}$ for both wavelengths.
We calibrated the Ca and G-band images by dark-current subtraction and flat fielding using the {\tt fg\_prep} procedure in the {\it SSW} package.
Then we reduced spatial fluctuations by cross-correlating two consecutive images in the respective wavelengths.

The Spectro-Polarimetor \citep[SP; ][]{lites13} obtained the full polarization states (Stokes-I, Q, U, and V) in the \ion{Fe}{1} lines (6301.5 and 6302.5 \AA) with a sampling of $21.6$ m\AA.
The data were observed using the Fast Mapping mode observation, in which the spatial sampling was $0.32^{\prime\prime}$/pixel and the spatial resolution was approximately $0.3^{\prime\prime}$ along the slit.
The SP scanning was performed from 00:11 UT to 00:43 UT on March 15, 2015, and covered on FOV of $164^{\prime\prime} \times 164^{\prime\prime}$.
We calibrated the SP scan data using the {\tt sp\_prep} procedure in the {\it SSW} package \citep{lites_ichimoto_13}, and vector magnetic fields were derived from the calibrated Stokes profiles on the assumption of the Milne-Eddington atmosphere adopting the inversion code MIKSY (Yokoyama et al., private communication).
The $180^{\circ}$ ambiguity was resolved using the minimum-energy code ``{\tt ME0}'' \citep{leka09a, leka09b, leka12} based on the minimum energy method of \citet{metcalf94}.
Then we spatially co-aligned the SP vector magnetogram and Ca-line images by cross-correlating the Stokes-I and G-band images.
From the co-alignment images, we investigated the spatial and temporal correlation between chromospheric brightenings and photospheric magnetic field.

\subsection{Nonlinear Force-Free Field Modeling} \label{sec:analysis_nlfff}

We briefly review the modeling of the 3D coronal magnetic field using an NLFFF approximation \citep[e.g.,][]{wiegelmann_sakurai_12}.
We employ the MHD relaxation method developed in \citet{inoue14, inoue16}.
The bottom boundary is fixed with the vector magnetic fields ($B_x, B_y, B_z$) in the photosphere, observed at 00 UT on March 15, 2015, by {\it SDO}/HMI.
We then solve the MHD equations in 3D space starting with a potential field which is extrapolated from the $B_z$ component in the photosphere \citep{sakurai82}.
Note that pressure  and gravity are neglected here because the low $\beta$ approximation is well satisfied in the solar corona 
\citep{gary01}.
Therefore, we solve the MHD equations simplified in the zero $\beta$ approximation. 
The basic equations are as follows, 
   \begin{equation}
   \rho = |{\bf B}|,
   \end{equation}
    \begin{equation}
    \frac{\partial {\bf v}}{\partial t} 
                         = - ({\bf v}\cdot{\bf \nabla}){\bf v}
                           + \frac{1}{\rho} {\bf J}\times{\bf B}
                           + \nu{\bf \nabla}^{2}{\bf v},
   \label{eq_of_mo}    
   \end{equation}
  \begin{equation}
  \frac{\partial {\bf B}}{\partial t} 
                        =  {\bf \nabla}\times({\bf v}\times{\bf B}
                        -  \eta{\bf J})
                        -  {\bf \nabla}\phi, 
  \label{in_eq}
  \end{equation}
  \begin{equation}
  {\bf J} = {\bf \nabla}\times{\bf B},
  \label{Am_low}
  \end{equation}
  \begin{equation}
  \frac{\partial \phi}{\partial t} + c^2_{h}{\bf \nabla}\cdot{\bf B} 
    = -\frac{c^2_{h}}{c^2_{p}}\phi,
  \label{div_eq}
  \end{equation}
\noindent where $\rho$ is plasma density, ${\bf B}$ is the magnetic flux density, ${\bf v}$ is the velocity, ${\bf J}$ is the electric current density, and $\phi$ is a convenient potential to remove errors derived from ${\bf \nabla}\cdot {\bf B}$ \citep{dedner02}, respectively.
The pseudo density is assumed to be proportional to $|{\bf B}|$ in order to ease the relaxation by equalizing the Alfven speed in space.
$\nu$ is a viscosity fixed as $1.0 \times 10^{-3}$, and the coefficients $c_h^2$, $c_p^2$ in Equation (\ref{div_eq}) are also fixed with the constant values, 0.04 and 0.1, respectively.
The initial density condition is given by $\rho = |{\bf B}|$ and velocity is set to zero in all space.
The resistivity is given by $ \eta = \eta_0 + \eta_1 |{\bf J}\times{\bf B}||{\bf v}|/|{\bf B}|$ where $\eta_0 = 5.0\times 10^{-5}$ and $\eta_1=1.0\times 10^{-3}$ in non-dimensional units.
The second term is introduced to accelerate relaxation to the force free field, particularly in weak field regions.
At the boundaries, all of velocities are fixed to zero and the magnetic fields are fixed to potential fields except at the bottom boundary.
Details are described in \cite{inoue14}.
  
The numerical box covers $216 \times 216 \times 151.2 (Mm^3)$ which is given as $1.0 \times 1.0 \times 0.7$ in non-dimensional units.
The region is divided into $300 \times 300 \times 210$ grids which is a result of the $2 \times 2$ binning process of the original vector magnetic fields in the photosphere taken by {\it SDO}/HMI.

\section{Results} \label{sec:results}

\subsection{Overview of the C-class Flares and Filament Eruption} \label{sec:results_overview}

\begin{figure}[ht!]
\plotone{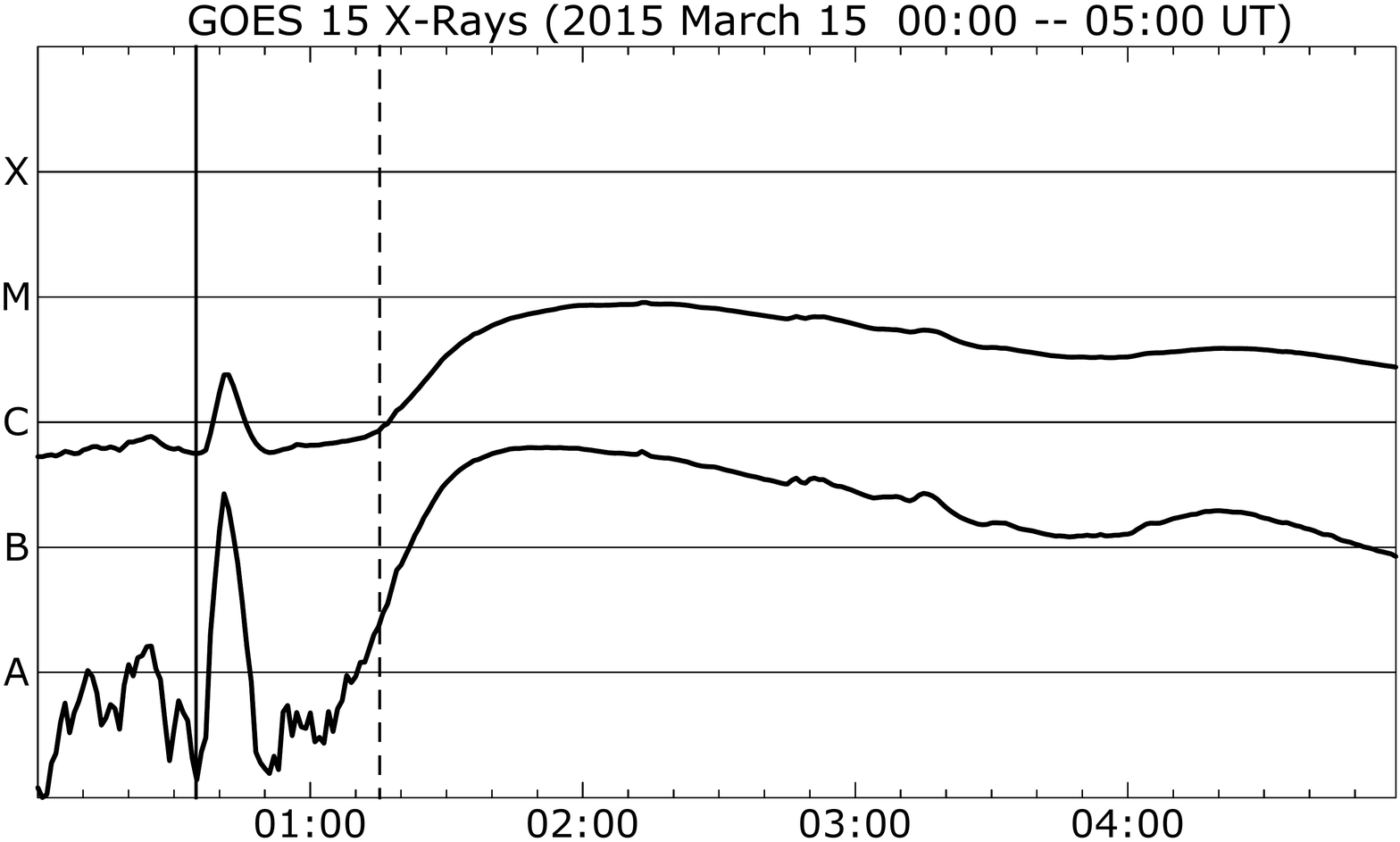}
\caption{
{\it GOES} soft X-ray light curve (1-8 {\AA} and 0.5-4 {\AA}) for 00-05 UT on March 15, 2015.
The vertical solid and broken lines indicate the onset times of the C2.4 and C9.1 flares (00:34 UT and 01:15 UT), respectively.
}
\label{fig:goes}
\end{figure}

\begin{figure}[ht!]
\plotone{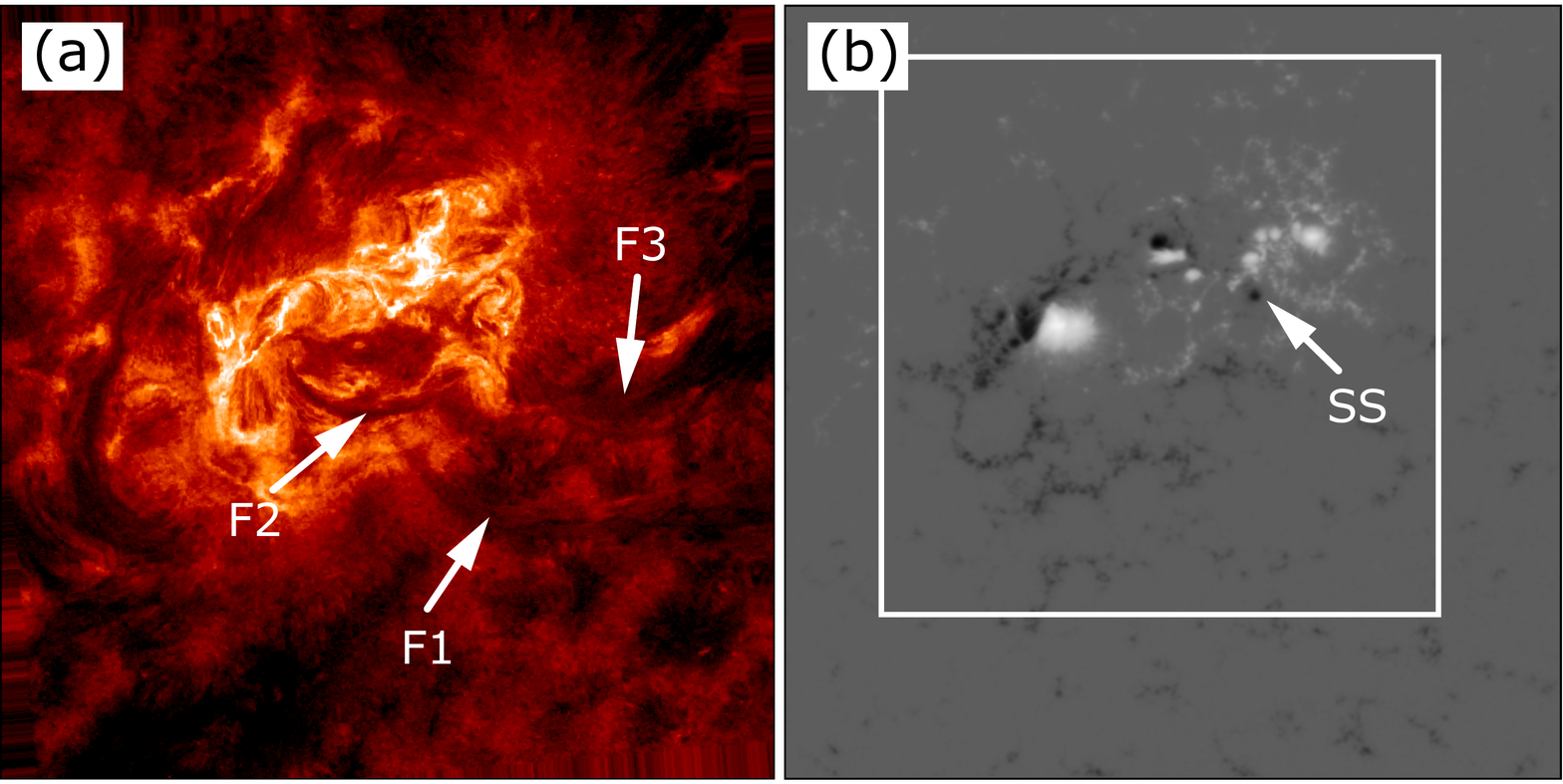}
\caption{
(a) Chromospheric filament structures in the perimeter region of AR 12297 obtained by {\it SDO/AIA} 304 \AA at 00 UT on March 15, 2015.
The spatial size of the FOV is $0.03^{\circ} \times 0.03^{\circ}$ in heliocentric angles, and the central coordinate is $195^{\circ}$ Carrington Longitude and $20^{\circ}$ latitude south.
Three filaments are labeled as F1, F2, and F3. F1 was the erupting filament associated with the C-class flares on March 15 while the other two did not erupt.
(b) Radial component of the photospheric magnetic field obtained by {\it SDO}/HMI.
The observation time and FOV are the same as the AIA image in panel (a).
The C2.4 flare occurred northeast of the PIL of a satellite spot (SS) that is indicated by the white arrow.
The white rectangle shows the FOV of the magnetogram in Figure~\ref{fig:twist_current}(a).
}
\label{fig:hmiaia}
\end{figure}

First, we review the evolution steps from the C2.4 flare to the huge filament eruption, using {\it SDO} data.
Figure~\ref{fig:goes} shows the {\it GOES} soft X-ray light curve covering the time of the C2.4 and C9.1 flares.
The onset time of the C2.4/C9.1 flare was 00:34 UT/01:15 UT on March 15, 2015, as indicated by the vertical solid/broken lines.
There were multiple filament structures close by AR 12297 as seen in the {\it SDO}/AIA 304 \AA image of Figure~\ref{fig:hmiaia}(a).
The erupting filament F1 and the other two filaments F2 and F3 are all rooted in the AR.
Panel (b) shows the radial component of the {\it SDO}/HMI magnetogram with the same FOV as panel (a).
The C9.1 flare occurred in association with the eruption of F1, and was a long duration event (LDE), while the preceding C2.4 flare was a relatively impulsive event as seen in Figure~\ref{fig:goes}.
The C2.4 flare occurred in the northern PIL of a satellite spot (SS) which is highlighted by the white arrow in panel (b).

The temporal variation of filaments F1, F2, and F3 related to the C2.4 and C9.1 flares is shown in the AIA 304 {\AA} images of Figure~\ref{fig:evolution} and Movie 1\footnote{Movie 1 is available in electronic article.}.
Panel (a) is a snapshot before the C2.4 flare.
There is a small filament in addition to the large filaments F1, F2, and F3 as indicated by the white arrow.
The small filament exists on the PIL which is located in the north of the SS in Figure~\ref{fig:hmiaia}(b).
The flare ribbons of the C2.4 flare appear as indicated by the black arrows in panel (b), and the southern ribbon is located over the PIL where the small filament is located.
The small filament erupts around 00:38 UT in association with the C2.4 flare as illustrated by the blue arrow \textcolor{red}{in panel (c)}.
\citet{wang16} investigated the detailed temporal evolution of the small filament eruption using AIA 304 {\AA} and 171 {\AA} images and {\it RHESSI}\footnote{{\it Reuven Ramaty High Energy Solar Spectroscopic Imager}}.
According to their analysis, the small filament eruption is composed of a hot X-ray jet and surge-like eruption and is analogous to the process proposed by \citet{sterling15}.
We found that F1 is drastically disturbed from 00:40 UT and it erupted southwestward (panel (d)) while the small filament eruption attenuate.
The two flare ribbons of the C9.1 flare gradually appeared from 01:12 UT corresponding to the huge F1 eruption as indicated by the black arrows in panel (e).
The ribbons and post flare arcade remained for more than 6 hours as the {\it GOES} soft X-ray light curve in Figure~\ref{fig:goes} gradually decayed.
Incidentally, F3 dynamically rose upward corresponding to the F1 eruption and showed a helical loop structure as indicated by the black arrow in panel (f), but it failed to erupt.
F2 stayed static and did not erupt with the C2.4 or 9.1 flares or the F1 eruption.

\subsection{Detailed Observation of the C2.4 Flare} \label{sec:results_obs}

It is inferred from the overview of the {\it SDO} analysis in the previous section (Section~\ref{sec:results_overview}) that the onset of the C2.4, C9.1, and the huge filament eruption is sequential.
Hence we study the details of the C2.4 flare, that seems to be related to the cause that triggers the sequential flares and eruption, using the {\it Hinode} data.
Figure~\ref{fig:hinode}(a, b) are chromospheric images before and just after the C2.4 flare onset, respectively.
A tiny brightening that is labeled PB by the white arrow in panel (a) is the last precursor brightening that was observed in the 20 minutes period from 23:10 UT (i.e. from 24 minutes before the C2.4 flare onset).
Then the initial flare ribbons of the C2.4 flare appeared as in panel (b); the separated two positive ribbons are labeled PR1 and PR2, and the negative ribbon is labeled NR.
The ribbons are also seen in Figure~\ref{fig:evolution}(b) but PR1 and NR are now clearly identified thanks to the high spatial resolution of {\it Hinode}.
The red cross in panel (b) marks the location of PB in panel (a).
The ribbons PR1, PR2, and NR were distributed with a circular shape on both sides of the location where PB was observed.

The background grayscale in panel (c) indicates the opposite polarities in the radial component of the photospheric magnetic field.
The green lines show the PILs and white/black arrows indicate the field strength and orientation of the horizontal magnetic fields.
PB was seen on the PIL located in the north of the SS, and the location is marked by the red cross.
We calculated the magnetic shear angles relative to the potential field in this region and show the distribution of the angles in panel (d).
Red and blue indicate the shear angles in the range ${\pm}$ 180$^{\circ}$.
The black lines and green cross show the PIL and location of PB, which are the same as in panel (c).
Obviously there are highly sheared fields (approximately 60$^{\circ}$ to 90$^{\circ}$ clockwise) along the flaring PIL on which the precursor brightening was seen.

\subsection{Magnetic Topology Related to Erupting Filament and C2.4 Flare} \label{sec:results_nlfff}

\begin{figure}[ht!]
\plotone{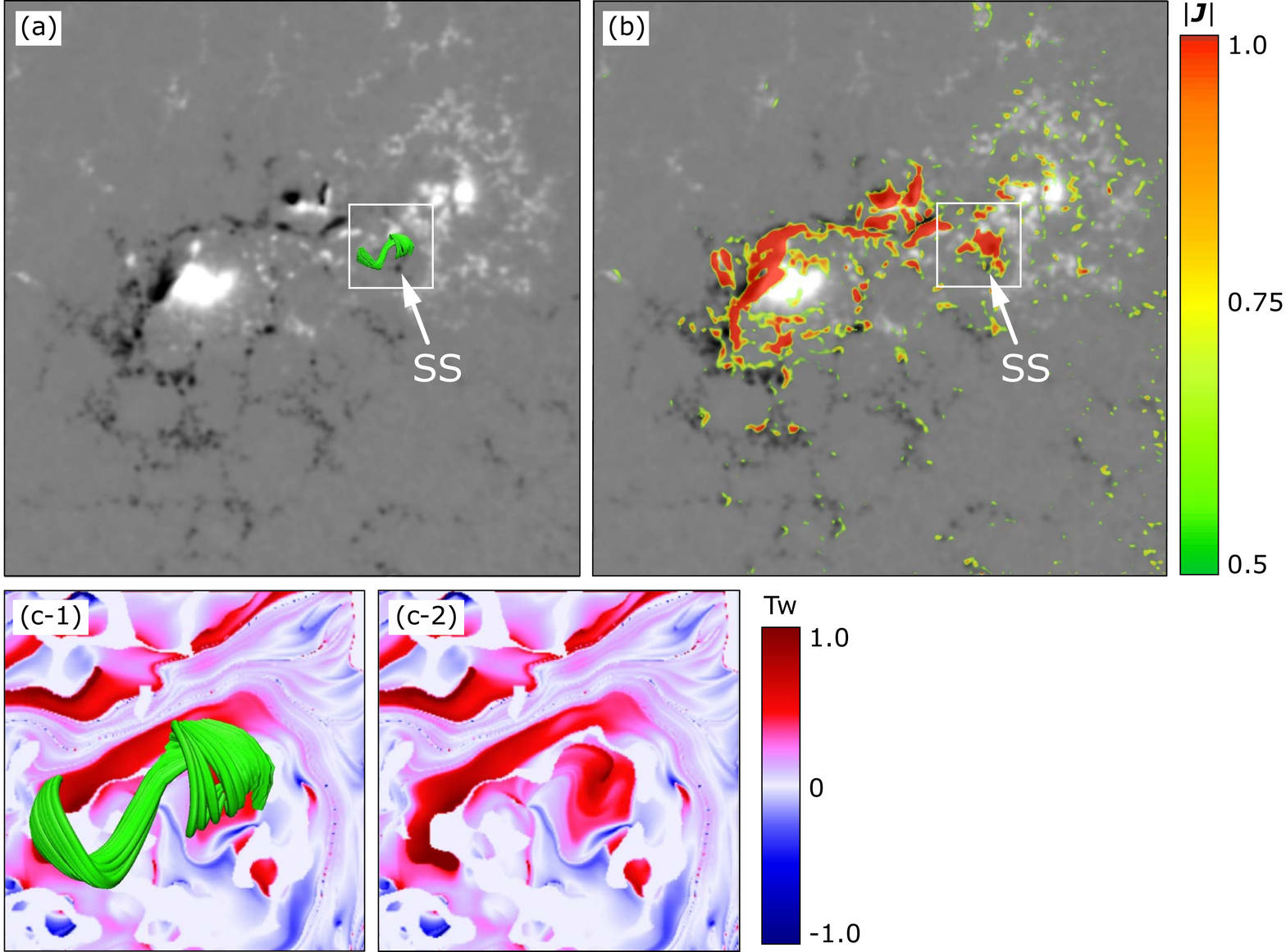}
\caption{
(a) Magnetic field lines extrapolated using the NLFFF method (green tubes). The background white/black indicates positive/negative component of the radial magnetic field in the range ${\pm}$ 2000 G.
(b) Spatial distribution of the electric current density $|\bm{J}|$. The background grayscale is identical to that in panel (a) and red color indicates the existence of strong electric currents.
(c) Spatial distribution of strong magnetic twist around the C2.4 flaring PIL, which is surrounded by the white square in panels (a, b). The magnetic twist is represented by the blue to red colors in the range $-1.0 < T_{W} < 1.0$. The green colored magnetic field lines of panel (a) are overlaid on the twist map in panel (c-1) while panel (c-2) shows only the twist distribution in the same region.
}
\label{fig:twist_current}
\end{figure}

We derived the coronal magnetic field using the NLFFF extrapolation around the C2.4 flaring PIL, where the precursor brightening was seen and magnetic shear was strong as seen in Figure~\ref{fig:hinode}.
Figure~\ref{fig:twist_current} shows the coronal magnetic field, the distribution of the electric current density $|\bm{J}|$, and the magnetic twist $T_{W}$ estimated from $1/4\pi\int\nabla\times\vec{B}\cdot\vec{B}/|\vec{B}^2|dl$, where $dl$ is a line element.
There is locally but strongly twisted field (green tubes), which is located north of the SS and is rooted at both sides of the C2.4 flaring PIL inside the white rectangle in panel (a).
Strong current concentrates on the C2.4 flaring region, although the electric current over the major PIL in the following spot dominates on the whole region, as can be seen in panel (b).
The magnetic twist in the C2.4 flaring region, marked by the white rectangles in panels (a, b), is seen in panels (c-1, c-2).
The red colored strong twist ($T_{W} \sim 1.0$) is distributed on both sides of the C2.4 flaring PIL, i.e., at the foot points of the local twisted magnetic field (green tubes).
Therefore, the northern PIL of the SS has locally highly twisted field with strong electric currents, and was more likely to flare.

We next investigated the relationship between the C2.4 flaring site and C9.1 erupting filament F1 on large-scales.
The green tubes in Figure~\ref{fig:aia_potential} show the closed magnetic field overlying F1.
The background images are AIA 304 {\AA} (panel a) and the radial component of the SHARP vector magnetogram in panel (b), respectively.
The three filament structures that are labeled in Figures~\ref{fig:hmiaia} and \ref{fig:evolution} are marked by the white arrows in panel (a).
The green-colored closed field lines are overlying F1, i.e., the closed fields prohibit F1 from erupting.
Panel (c) is a bird's-eye view from the direction of the blue arrow in panel (b).
It shows that one side of the foot points of the closed fields are rooted in the positive polarity region to the north of the C2.4 flaring region.
The foot point is close to the highly twisted compact field over the C2.4 flaring PIL shown in Figure~\ref{fig:twist_current}(a).
The other foot point of the closed fields is connected with both the SS and the weak negative polarity region in the south of the AR.
Moreover, we displayed in panel (d) a vertical cross-section of the decay index $n$ that is located on the dashed line in panel (b).
The decay index $n$ is defined as $n = -{d\ln{\mid{\bm B_{p, h}}\mid/}d\ln{\mid{z}\mid}}$ \citep{kliemtorok06}.
Here, ${\bm B_{p, h}}$ and $z$ are the horizontal component of the potential magnetic field calculated from the observed $B_{z}$ (radial component of HMI SHARP data) and the height, respectively.
Obviously, only a part of the closed fields reaches the isosurface of $n = 1.5$, which is indicated by the black line in panel (d) and corresponds to the critical value for the torus instability \citep[e.g.][]{zuccarello15}.
This suggests that the filament that is trapped by the closed field is stable with respect to the torus instability.

\subsection{Behavior of the Erupting Filament} \label{sec:results_filament}

We investigated the temporal variation of the erupting filament F1 through 24 hours to the C9.1 eruption onset.
Figure~\ref{fig:193slice} shows time-slice plots of F1 in {\it SDO}/AIA 193 {\AA}.
We cut the erupting filament along the white vertical line in panel (a), which is roughly along the green-colored closed field in Figure~\ref{fig:aia_potential}.
The time-slice plots along the line-cut in (a) are panels (b-1) and (b-2), respectively.

\begin{figure}[ht!]
\plotone{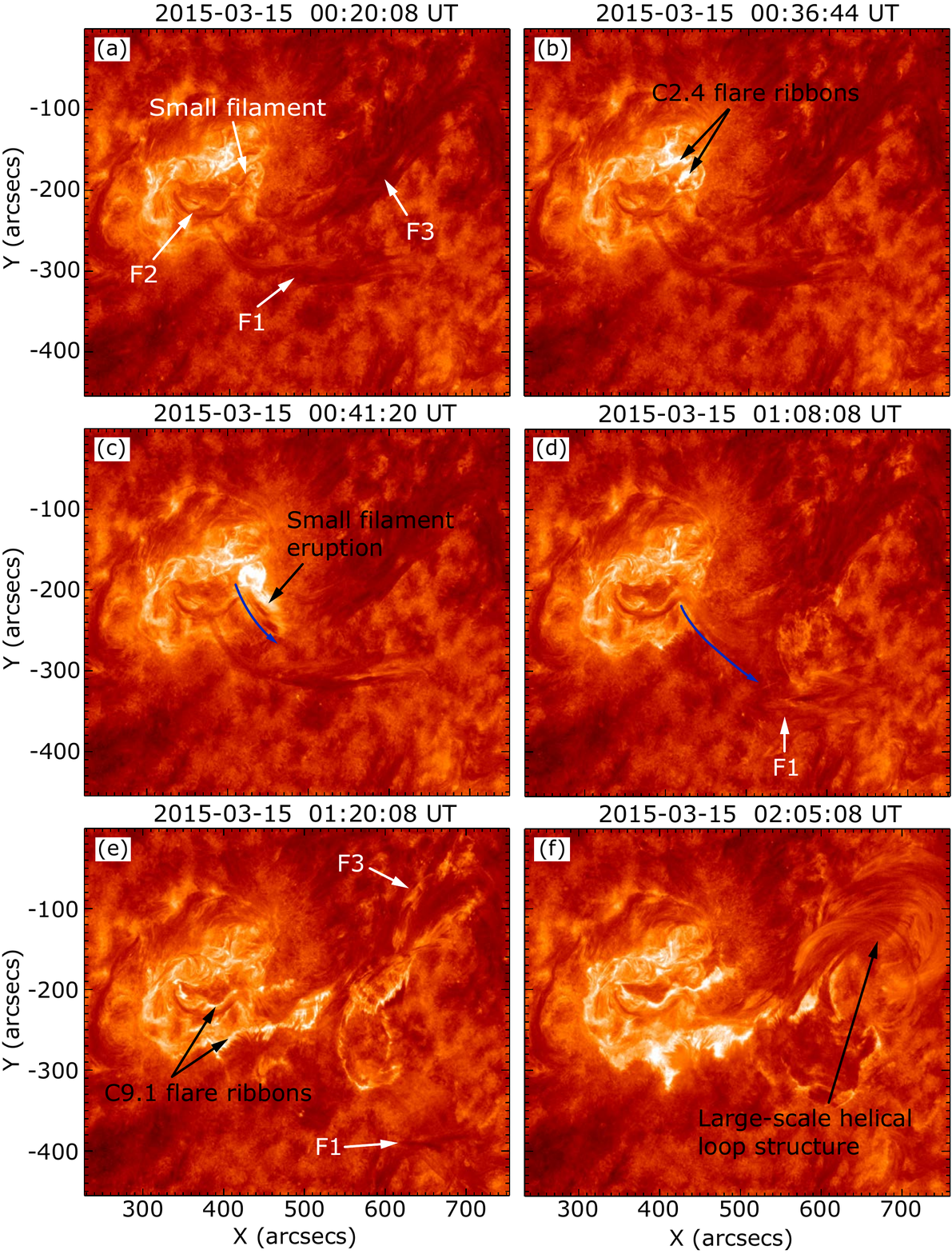}
\caption{
Temporal evolution of the filaments and flare ribbons before and after the C2.4 flare onset.
(a) Multiple filament structure (F1, F2, F3) and a small filament that exists over the northern PIL of SS.
(b) Flare ribbons associated with the C2.4 flare.
(c, d) A snapshot of the small filament eruption. The erupting direction is represented by the blue arrows.
(e) Flare ribbons that appeared in association with the F1 eruption and the C9.1 flare.
(f) Large-scale helical loop structure that appeared related to the F3 disturbance (failed eruption).
The corresponding Movie 1 is available in electronic manuscript.
}
\label{fig:evolution}
\end{figure}

Panel (b-1) shows 24 hours of the evolution of F1.
F1, which is marked by the white arrow, gradually moves downward in the image (southward in panel (a)) in the 12 hour period from 00 UT on March 14.
Note that downward movement in the image means movement away from the AR on the Sun because the AR and filaments are located in the southwest hemisphere.
A bright feature around 12 UT on March 14 is related to an eruption of F2 with a C2.6 flare (11:44 UT onset) followed by a first CME \citep{wang16}.
According to the $H_{\alpha}$ observation in \citet{chandra17}, the filaments F1 and F3, which are labeled in Figure~\ref{fig:evolution}(a) in the present paper, were originally one huge filament.
This F2 eruption disturbed the huge filament, and the huge filament appeared divided into F1 and F3, as can be seen in Figure~\ref{fig:evolution}(a).
The disturbances from F1 last until around 18:30 UT, then F1 moves downward in a step-like motion as can be seen in panel (b-1).

Panel (b-2) focus onto the time period from 20 UT on March 14 to 03 UT on March 15, i.e., from 4.5 hours before the C2.4 flare onset to after the C9.1 eruption.
The onset times of the C2.4 and C9.1 flares are indicated by the blue and green vertical lines, respectively.
The white vertical broken lines indicate the start and end times of the precursor brightening, which is labeled as PB in Figure~\ref{fig:hinode}.
It is noteworthy that F1 move downward very slowly but become unstable just after the PB was observed.
The C2.4 flare occurred following the PB and F1 disturbance, then F1 erupted.

\section{Discussion} \label{sec:discussion}

The time series of observational facts derived from our data analysis is summarized below.
\begin{enumerate}
	\item The closed field that roots the positive polarity region in the north of the SS holds filament F1 while there is compact but highly twisted field over the north PIL of the SS.
	\item The tiny brightening PB appeared on the north PIL of the SS, then the filament F1 is disturbed afterwards.
	\item The C2.4 flare, which shows circular shaped ribbons, occurs and the small filament that existed on the northern PIL of SS erupts.
	\item The filament F1 is drastically disturbed and erupts. Huge C9.1 flare ribbons are observed in the south of the AR and gradually enhance.
\end{enumerate}
The F1 eruption related to the C9.1 event was accompanied by a fast CME that caused the largest magnetic storm in solar cycle 24.
Here, we discuss the onset process of the huge geo-effective solar eruption.

\begin{figure}[ht!]
\plotone{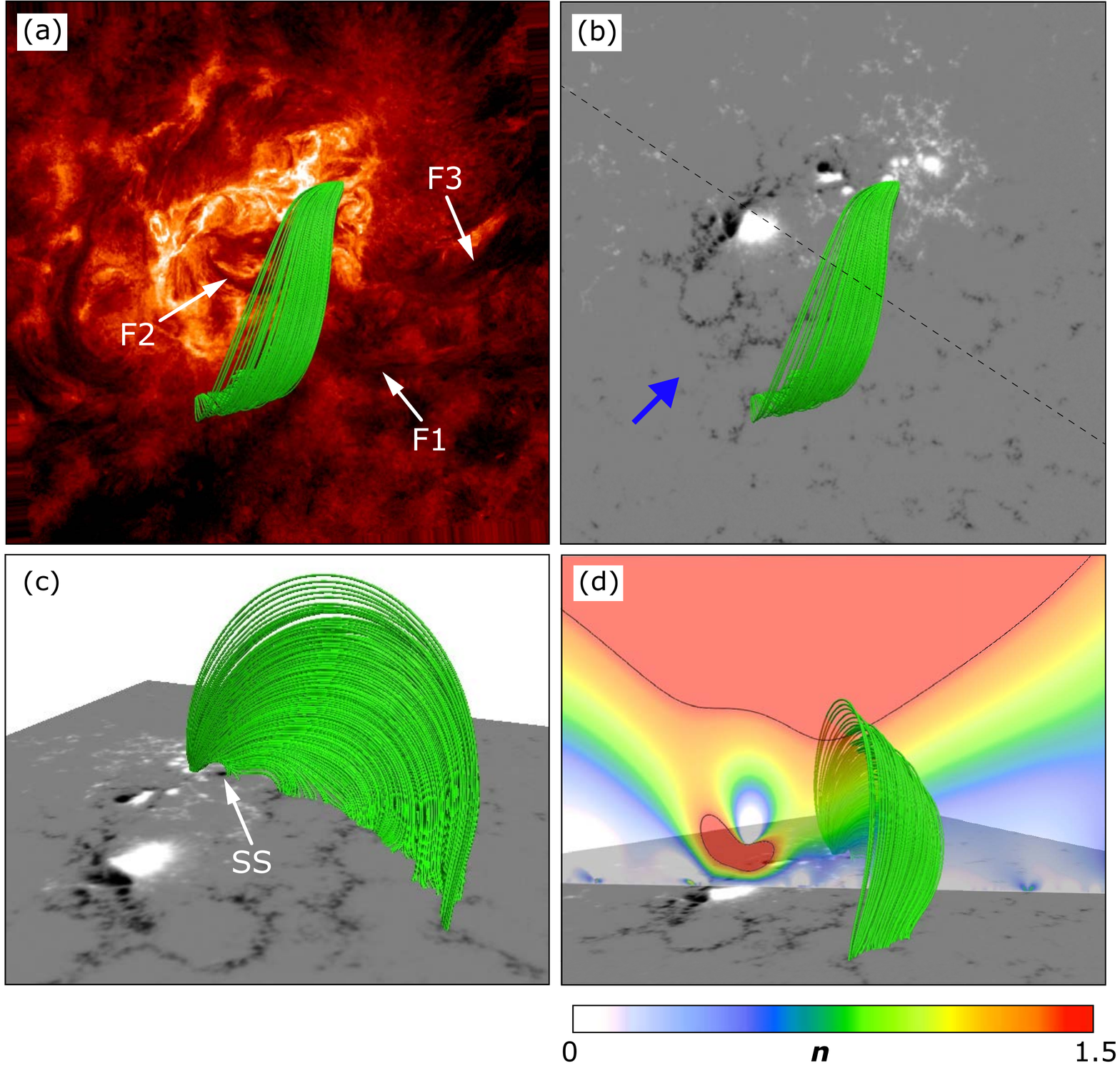}
\caption{
Magnetic field lines that prohibit F1 from eruption.
The background is an AIA 304 {\AA} image in panel (a) and the radial component of the magnetic field in panel (b) at 00 UT on March 15.
The bird's-eye view from the direction that is indicated by the blue arrow in panel (b) is shown in panel (c).
The green tubes in all the panels are the potential field that roots in the C2.4 flaring region.
The vertical cross-section in panel (d) displays the decay index $n$ between 0 and 1.5 located on the dashed line in panel (b).
The decay index $n$ is calculated form the horizontal component of the potential field extrapolated from the vector magnetic field.
The black line on the vertical cross-section indicates $n = 1.5$, which corresponds to the critical value for the torus instability.
}
\label{fig:aia_potential}
\end{figure}

The torus instability is one candidate to explain an initiation of the CME.
Note that, in order to drive the instability,  the axis of the magnetic flux needs to reach a region where the decay index value satisfies $n > 1.5$ \citep{zuccarello16}.
Therefore, other driving processes (e.g., kink instability, duble-arc sustainability \citep{ishiguro_kusano17}) are also important to push the flux rope to a region where it becomes unstable to the torus instability.
In our NLFFF modeling result in Figure~\ref{fig:aia_potential}, the flux rope sustaining filament F1 is inferred to be stable to the torus instability, at least, at 00 UT on March 15 because the green colored field lines strapping down the flux rope pass through a region where the decay index value satisfies $n \sim 1.5$.
On the other hand, F1 gradually rises upward as shown in Figure~\ref{fig:193slice}, which converts to a quick ascension just after the C2.4 flare.
Therefore, the C2.4 flare might accelerate the instability of F1.
Figure~\ref{fig:aia_potential} shows that the C2.4 flare was observed in the region where the closed field lines in green, which strap down F1, are rooted.
Therefore the C2.4 flare might play a role in disturbing and initiating the eruption of F1.
Several studies \citep[e.g.][]{olmedo_zhang10, myers15} reported that the critical decay index value strongly depends on the shape or boundary state of a flux rope, consequently, the flux rope becomes unstable to the torus instability even though its axis does not reach a region where $n > 1.5$ is satisfied.
Therefore, the C2.4 flare might accelerate the torus instability.

\begin{figure}[ht!]
\plotone{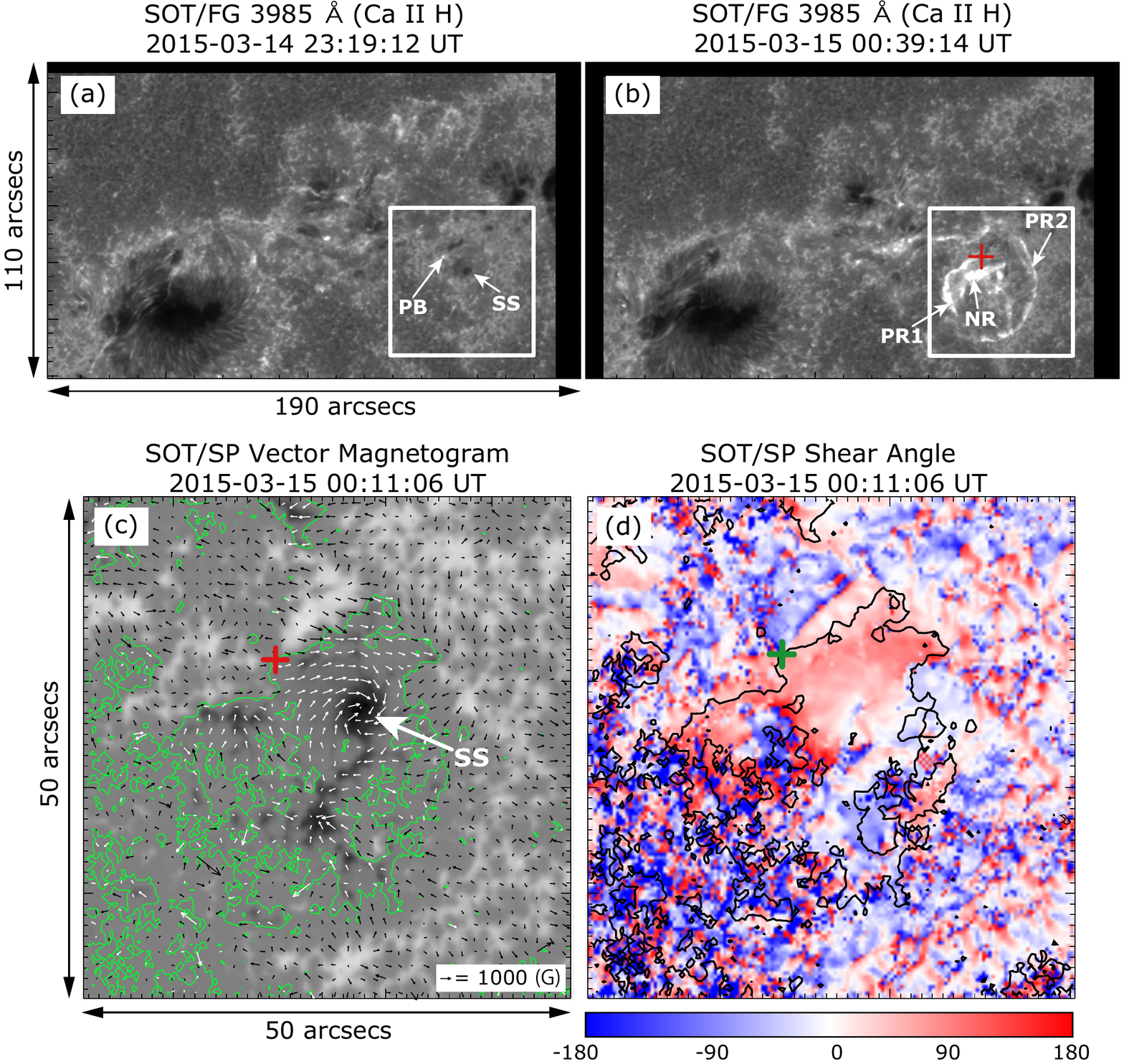}
\caption{
(a, b) Chromospheric images obtained by {\it Hinode}/SOT in the \ion{Ca}{2} H line.
The satellite spot SS that is seen in Figure~\ref{fig:hmiaia}(b) is pointed out by the white arrow in panel (a) (also in panel (c)). 
The precursor brightening (PB) and initial flare ribbons of the C2.4 flare (positive ribbons: PR1 and PR2, and negative ribbon) are indicated by the white arrows.
The red cross mark in panel (b) indicates the location of the PB that is labeled in panel (a).
(c) Photospheric magnetic field in the C2.4 flaring region surrounded by the white rectangles in panels (a, b), observed by SP.
The background white/black indicates positive/negative of the radial components of the magnetic field in the range ${\pm}$ 2000 G.
Green lines are PILs and white/black arrows are horizontal components larger than 100 G.
The red cross marks the location of the PB seen in panel (a).
(d) Distribution map of the magnetic shear angle of the horizontal field relative to the potential field in the range ${\pm}$ 180$^{\circ}$.
Red/blue color indicates clockwise/counter-clockwise shear.
Black lines and the green cross indicate PILs and the location of the PB.
}
\label{fig:hinode}
\end{figure}

\begin{figure}[ht!]
\plotone{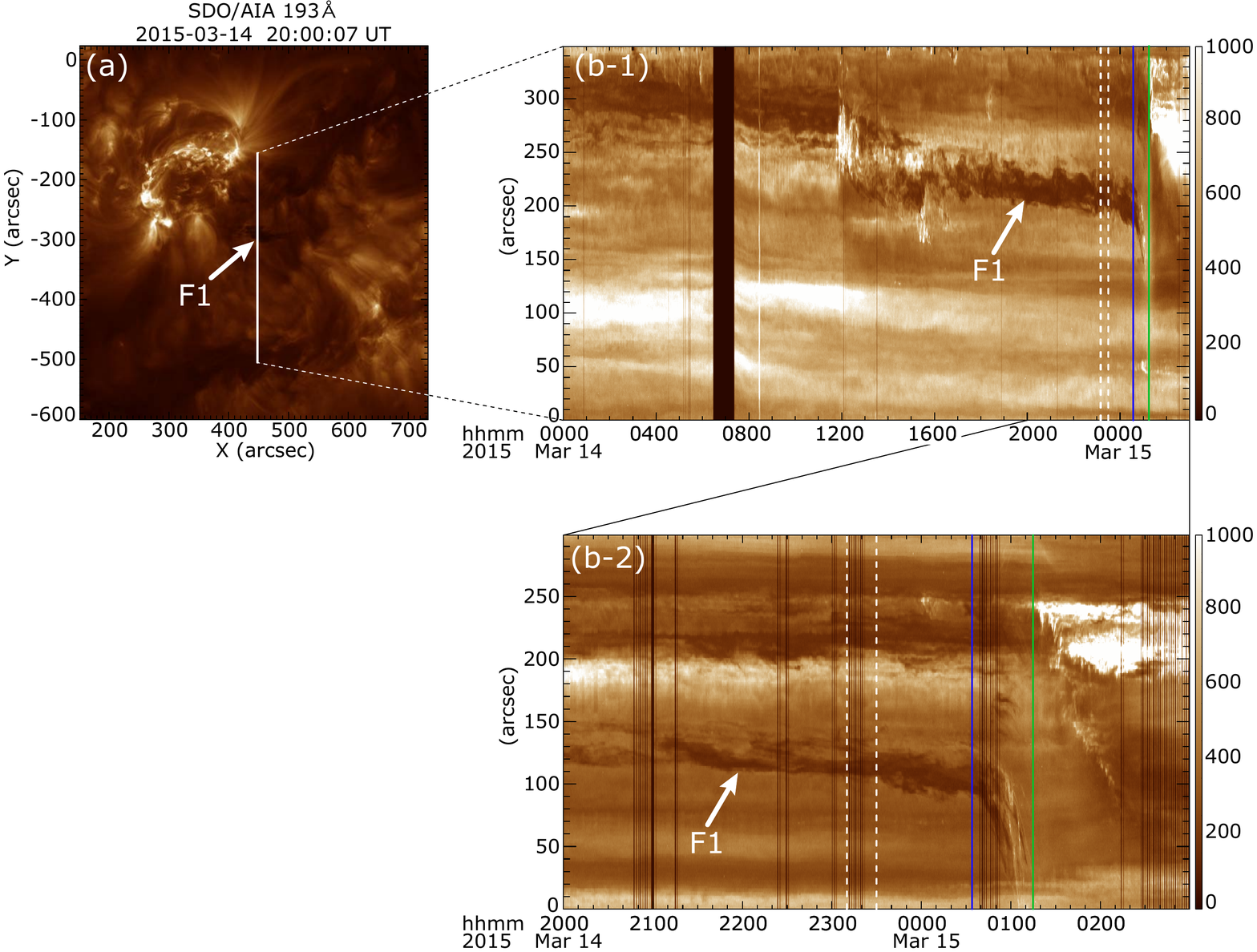}
\caption{
Temporal variation of the erupting filament observed by {\it SDO}/AIA 193 {\AA}.
(a) The erupting filament F1 is marked by the white arrow. The line-cut is over plotted by the white line.
(b-1, b-2) Time-slice image of F1. The C2.4 and C9.1 flaring times are indicated by the vertical blue and green lines, respectively. The duration of the small precursor brightening (PB in Figure~\ref{fig:hinode}) is indicated as between the vertical white dashed lines. The upper plot (b-1) shows the day-order variation from 00 UT on March 14 to 03 UT on March 15, 2015. The bottom plot (b-2) shows the detailed temporal variation of F1 from 20:00 UT on March 14 to 03 UT on March 15, 2015. 
}
\label{fig:193slice}
\end{figure}

We further propose that the precursor brightening (PB) that was seen just before the C2.4 flare is possibly related to a trigger of the F1 eruption.
This is inferred from the fact that the precursor brightening PB was observed on the C2.4 flaring PIL located between the C2.4 flare ribbons, and become F1 was disturbed from just after PB was observed (Figures~\ref{fig:hinode} and \ref{fig:193slice}).
Thus, it is inferred that a small-scale magnetic field disturbance caused the PB in Figure~\ref{fig:hinode}.
This disturbance triggered magnetic reconnection of the locally twisted field seen in Figure~\ref{fig:twist_current} and the C2.4 flare occurred in association with the small filament eruption of Figure~\ref{fig:evolution} (b) \citep[e.g.][]{chenshibata00, Green11, bamba13, sterling15, wang16}.
From these results, we suggest that emerging magnetic flux \citep{chenshibata00} or magnetic flux cancellation \citep{Green11} cause the tiny precursor brightening PB.
It is difficult to identify what magnetic disturbance is suitable to cause the tiny brightening PB because the spatial scale is smaller than the spatial resolution of {\it SDO}/HMI and AIA.
However, our analysis results support the scenario that the sequential huge eruption started from a tiny brightening at the foot point independent from the C9.1 eruption.

\section{Summary} \label{sec:summary}

In this study, we aimed to understand the onset mechanism of the huge filament eruption which caused the largest magnetic storm, the so-called ``St Patrick's day storm'', in solar cycle 24.
We focused on the C2.4 flare that occurred approximately 45 minutes prior to the C9.1 flare onset corresponding to the huge filament eruption.
Using {\it Hinode}/SOT, {\it SDO}/HMI and AIA, and NLFFF modeling, we investigated the behavior of the erupting filament and characteristics of the photospheric/coronal magnetic field and chromospheric brightening before/after the C2.4 flare.

We found a tiny precursor brightening that was observed on the PIL at the C2.4 flaring site.
There was highly twisted magnetic field structure locally in the region over the precursor brightening which is seen as a small filament.
The C2.4 flaring site, where the twisted field (i.e. the small filament) and the precursor brightening were observed, was a foot point of the closed magnetic field that sustained the erupting filament.
From a comparison between the observational results and NLFFF modeling, it is suggested that the tiny precursor brightening is the proxy of a small-scale magnetic disturbance that triggered the sequential eruption from the small filament to the filament that caused the geomagnetic storm.

It would be no surprise if the erupting filament had erupted at some point without any ``trigger'', because it gradually rose upward from at least 24 hours before the C9.1 eruption onset.
However, the filament was stable with respect to the torus mode at least until 00 UT on March 15, just before the C2.4 flare onset according to our NLFFF modeling.
Our observational results showed us that the erupting filament is destabilized after the precursor brightening was observed, then it erupts subsequent to the C2.4 flare and small filament eruption.

Therefore, in the present study, we suggest that the tiny magnetic disturbance, which caused the precursor brightening, the C2.4 flare, and accompanying small filament eruption, destabilized the closed field that prohibited the filament F1 from erupting.
This triggers the huge filament eruption which was the source of the largest magnetic storm in solar cycle 24, the so-called ``St Patrick's day storm''.


\acknowledgments

This study was triggered by the workshop on "Solar-Terrestrial Environment Prediction: Could we predict the storm on March 17?" held in Nagoya University on 2015 April 21 supported by the Project for Solar-Terrestrial Environment Prediction (PSTEP).
The authors deeply appreciate the various fruitful discussions and comments from the researchers who joined the workshop.
We further thank the researchers and students in ISAS/JAXA, ISEE/Nagoya University, NWRA, NAOHJ, and MSSL/UCL, for their valuable comments on our study.
Hinode is a Japanese mission developed and launched by ISAS/JAXA, which collaborates with NAOJ as a domestic partner and with NASA and STFC (UK) as international partners.
Scientific operation of the Hinode mission is conducted by the Hinode science team organized at ISAS/JAXA.
The HMI and AIA data have been used courtesy of NASA/{\it SDO} and the AIA and HMI science teams.
This work was partly carried out at the Hinode Science Center at NAOJ and Nagoya University, Japan, and the Solar Data Analysis System (SDAS) operated by the Astronomy Data Center in cooperation with the Hinode Science Center of the NAOJ.
The visualization was done using VAPOR \citep{clyne05, clyne07}.
The authors thank the MEXT/KAKENHI "Exploratory Challenge on Post-K Computer (Environmental Variations of Planets in Solar System)" for supporting this study.
This work was supported by MEXT/JSPS KAKENHI Grant Numbers JP15J10092, JP16H07478, JP15H05812, JP15H05814, and JP15K21709.
YB thank Dr. T. Hara for kind technical support on SPEDAS.
Data processing was done using SPEDAS V3.1, see \citet{angelopoulos19}.
The authors thank the anonymous referee for valuable comments that improved the clarity of the manuscript.

\vspace{5mm}
\facilities{{\it Hinode} (SOT), {\it SDO} (HMI and AIA)}
\software{SSW, SPEDAS, VAPOR}





\begin{thebibliography}{}

\bibitem[Angelopoulos et al.(2019)]{angelopoulos19} Angelopoulos, V., Cruce, P., Drozdov, A. et al. Space Sci Rev (2019) 215: 9. 
\bibitem[Antiochos et al.(1999)]{antiochos99} Antiochos, S.~K., DeVore, C.~R., \& Klimchuk, J.~A.\ 1999, \apj, 510, 485 
\bibitem[Bamba et al.(2013)]{bamba13} Bamba, Y., Kusano, K., Yamamoto, T.~T., \& Okamoto, T.~J.\ 2013, \apj, 778, 48
\bibitem[Bamba et al.(2017a)]{bamba17a} Bamba, Y., Inoue, S., Kusano, K., \& Shiota, D.\ 2017, \apj, 838, 134 
\bibitem[Bamba et al.(2017b)]{bamba17b} Bamba, Y., Lee, K.-S., Imada, S., \& Kusano, K.\ 2017, \apj, 840, 116
\bibitem[Bateman(1978)]{bateman78} Bateman, G.\ 1978, Cambridge, Mass., MIT Press, 1978.~270 p.
\bibitem[Bobra et al.(2014)]{bobra14} Bobra, M.~G., Sun, X., Hoeksema, J.~T., et al.\ 2014, \solphys, 289, 3549.
\bibitem[Chandra et al.(2017)]{chandra17} Chandra, R., Filippov, B., Joshi, R., \& Schmieder, B.\ 2017, \solphys, 292, 81 
\bibitem[Chen \& Shibata(2000)]{chenshibata00} Chen, P.~F., \& Shibata, K. 2000, {\it ApJ}, 545, 524
\bibitem[Clyne \& Rast(2005)]{clyne05} Clyne, J., \& Rast, M.\ 2005, \procspie, 5669, 284
\bibitem[Clyne et al.(2007)]{clyne07} Clyne, J., Mininni, P., Norton, A., \& Rast, M.\ 2007, New Journal of Physics, 9, 301
\bibitem[Dedner et al.(2002)]{dedner02} Dedner, A., Kemm, F., Kr{\"o}ner, D., et al.\ 2002, Journal of Computational Physics, 175, 645 
\bibitem[D{\'e}moulin \& Aulanier(2010)]{demoulin10} D{\'e}moulin, P., \& Aulanier, G.\ 2010, \apj, 718, 1388 
\bibitem[Dud{\'\i}k et al.(2017)]{dudik17} Dud{\'\i}k, J., Zuccarello, F.~P., Aulanier, G., et al.\ 2017, \apj, 844, 54.
\bibitem[Fan \& Gibson(2003)]{fangibson03} Fan, Y., \& Gibson, S.~E.\ 2003, {\it ApJL}, 589, L10
\bibitem[Gerrard et al.(2001)]{gerrard01} Gerrard, C.~L., Arber, T.~D., Hood, A.~W., \& Van der Linden, R.~A.~M.\ 2001, \aap, 373, 1089
\bibitem[Gary(2001)]{gary01} gary, G.~A.\ 2001, \solphys, 203, 71 
\bibitem[Green et al.(2011)]{Green11} Green, L.~M., Kliem, B., \& Wallace, A.~J.\ 2011, \aap, 526, A2 
\bibitem[Ishiguro, \& Kusano(2017)]{ishiguro_kusano17} Ishiguro, N., \& Kusano, K.\ 2017, \apj, 843, 101.
\bibitem[Inoue et al.(2014)]{inoue14} Inoue, S., Magara, T., Pandey, V.~S., et al.\ 2014, \apj, 780, 101 
\bibitem[Inoue et al.(2015)]{inoue15} Inoue, S., Hayashi, K., Magara, T., Choe, G.~S., \& Park, Y.~D.\ 2015, \apj, 803, 73 
\bibitem[Inoue(2016)]{inoue16} Inoue, S.\ 2016, Progress in Earth and Planetary Science, 3, 19 
\bibitem[Karpen et al.(2012)]{karpen12} Karpen, J.~T., Antiochos, S.~K., \& DeVore, C.~R.\ 2012, \apj, 760, 81 
\bibitem[Kataoka et al.(2015)]{kataoka15} Kataoka, R., Shiota, D., Kilpua, E., \& Keika, K.\ 2015, \grl, 42, 5155 
\bibitem[Kliem \& T{\"o}r{\"o}k(2006)]{kliemtorok06} Kliem, B., T{\"o}r{\"o}k, T.\ 2006, {\it Physical Review Letters}, 96, 255002 
\bibitem[Kliem et al.(2014)]{kliem14} Kliem, B., Lin, J., Forbes, T.~G., Priest, E.~R., \& T{\"o}r{\"o}k, T.\ 2014, \apj, 789, 46
\bibitem[Knipp et al.(2018)]{knipp18} Knipp, J. D., Fraser, J. B., Shea, A. M., \& Smart, F. D. \ 2018, Space Weather, 16, ***
\bibitem[Kosugi et al.(2007)]{kosugi07} Kosugi, T., Matsuzaki, K., Sakao, T., et al.\ 2007, {\it Solar Phys.}, 243, 3 
\bibitem[Labrosse et al.(2010)]{labrosse10} Labrosse, N., Heinzel, P., Vial, J.-C., et al.\ 2010, \ssr, 151, 243 
\bibitem[Leka et al.(2009a)]{leka09a} Leka, K.~D., Barnes, G., \& Crouch, A.\ 2009, The Second Hinode Science Meeting: Beyond Discovery-Toward Understanding, 415, 365 
\bibitem[Leka et al.(2009b)]{leka09b} Leka, K.~D., Barnes, G., Crouch, A.~D., et al.\ 2009, \solphys, 260, 83 
\bibitem[Leka et al.(2012)]{leka12} Leka, K.~D., Barnes, G., Gary, G.~A., Crouch, A.~D., \& Liu, Y.\ 2012, \solphys, 276, 441
\bibitem[Lemen et al.(2012)]{lemen12} Lemen, J.~R., Title, A.~M., Akin, D.~J., et al.\ 2012, \solphys, 275, 17
\bibitem[Lites et al.(2013)]{lites13} Lites, B.~W., Akin, D.~L., Card, G., et al.\ 2013, \solphys, 283, 579
\bibitem[Lites \& Ichimoto(2013)]{lites_ichimoto_13} Lites, B.~W., \& Ichimoto, K.\ 2013, \solphys, 283, 601 
\bibitem[Liu et al.(2015)]{liu15} Liu, Y.~D., Hu, H., Wang, R., et al.\ 2015, \apjl, 809, L34 
\bibitem[Louis et al.(2015)]{louis15} Louis, R.~E., Kliem, B., Ravindra, B., et al.\ 2015, \solphys, 290, 3641.
\bibitem[Mackay et al.(2010)]{mackay10} Mackay, D.~H., Karpen, J.~T., Ballester, J.~L., Schmieder, B., \& Aulanier, G.\ 2010, \ssr, 151, 333 
\bibitem[Masson et al.(2009)]{masson09} Masson, S., Pariat, E., Aulanier, G., \& Schrijver, C.~J.\ 2009, \apj, 700, 559 
\bibitem[Metcalf(1994)]{metcalf94} Metcalf, T.~R.\ 1994, \solphys, 155, 235
\bibitem[Moore et al.(2001)]{moore01} Moore, R.~L., Sterling, A.~C., Hudson, H.~S., \& Lemen, J.~R.\ 2001, {\it ApJ}, 552, 833
\bibitem[Moore \& Sterling(2006)]{moore06} Moore, R.~L., \& Sterling, A.~C.\ 2006, Washington DC American Geophysical Union Geophysical Monograph Series, 165, 43 
\bibitem[Myers et al.(2015)]{myers15} Myers, C.~E., Yamada, M., Ji, H., et al.\ 2015, \nat, 528, 526.
\bibitem[Olmedo, \& Zhang(2010)]{olmedo_zhang10} Olmedo, O., \& Zhang, J.\ 2010, \apj, 718, 433.
\bibitem[Pesnell et al.(2012)]{pesnell12} Pesnell, W.~D., Thompson, B.~J., \& Chamberlin, P.~C.\ 2012, \solphys, 275, 3
\bibitem[Sakurai(1982)]{sakurai82} Sakurai, T.\ 1982, \solphys, 76, 301.
\bibitem[Schou et al.(2012)]{schou12} Schou, J., Scherrer, P.~H., Bush, R.~I., et al.\ 2012, \solphys, 275, 229
\bibitem[Schrijver \& Title(2002)]{schrijver02} Schrijver, C.~J., \& Title, A.~M.\ 2002, \solphys, 207, 223 
\bibitem[Sterling et al.(2015)]{sterling15} Sterling, A.~C., Moore, R.~L., Falconer, D.~A., \& Adams, M.\ 2015, \nat, 523, 437 
\bibitem[T{\"o}r{\"o}k et al.(2004)]{torok04} T{\"o}r{\"o}k, T., Kliem, B., \& Titov, V.~S.\ 2004, \aap, 413, L27 
\bibitem[T{\"o}r{\"o}k \& Kliem(2005)]{torokkliem05}  T\"or\"ok, T., \& Kliem, B.\ 2005, {\it ApJ}, 630, L97
\bibitem[T{\"o}r{\"o}k, \& Kliem(2007)]{torokkliem07} T{\"o}r{\"o}k, T., \& Kliem, B.\ 2007, Astronomische Nachrichten, 328, 743.
\bibitem[Tsuneta et al.(2008)]{tsuneta08} Tsuneta, S., Ichimoto, K., Katsukawa, Y., et al.\ 2008, \solphys, 249, 167
\bibitem[Wang et al.(2015)]{wang15} Wang, H., Cao, W., Liu, C., et al.\ 2015, Nature Communications, 6, 7008.
\bibitem[Wang et al.(2016)]{wang16} Wang, R., Liu, Y.~D., Zimovets, I., et al.\ 2016, \apjl, 827, L12 
\bibitem[Wiegelmann \& Sakurai(2012)]{wiegelmann_sakurai_12} Wiegelmann, T., \& Sakurai, T.\ 2012, Living Reviews in Solar Physics, 9, 5 
\bibitem[Woods et al.(2017)]{Woods17} Woods, M.~M., Harra, L.~K., Matthews, S.~A., et al.\ 2017, \solphys, 292, 38 
\bibitem[Yan et al.(2017)]{yan17} Yan, X.~L., Jiang, C.~W., Xue, Z.~K., et al.\ 2017, \apj, 845, 18.
\bibitem[Yashiro et al.(2006)]{yashiro06} Yashiro, S., Akiyama, S., Gopalswamy, N., \& Howard, R.~A.\ 2006, \apjl, 650, L143
\bibitem[Xue et al.(2016)]{xue16} Xue, Z., Yan, X., Cheng, X., et al.\ 2016, Nature Communications, 7, 11837 
\bibitem[Zuccarello et al.(2015)]{zuccarello15} Zuccarello, F.~P., Aulanier, G., \& Gilchrist, S.~A.\ 2015, \apj, 814, 126 
\bibitem[Zuccarello et al.(2016)]{zuccarello16} Zuccarello, F.~P., Aulanier, G., \& Gilchrist, S.~A.\ 2016, \apjl, 821, L23 
\bibitem[Zuccarello et al.(2017)]{zuccarello17} Zuccarello, F.~P., Chandra, R., Schmieder, B., et al.\ 2017, \aap, 601, A26.

\end{thebibliography}
\end{document}